\documentclass[aps,prl,
superscriptaddress,
showpacs,
preprintnumbers,
nofootinbib,
amsmath,
amssymb,
aps,
twocolumn,
floatfix,
]{revtex4-1}

\usepackage{graphicx}
\usepackage{dcolumn}
\usepackage{bm}
\usepackage[usenames,dvipsnames]{color}
\usepackage{verbatim}
\usepackage{amsfonts}
\usepackage{geometry}
\usepackage{longtable}
\usepackage{natbib}
\usepackage{hyperref}
\usepackage[mathlines]{lineno}

\graphicspath{{figs/}}

\newcommand{\mres}{m_\text{res}}
\newcommand{\ua}{U(1)_A}

\newcommand{\sua}{SU(2)_L\times SU(2)_R}

\def\slashed#1{\kern+0.1em /\kern-0.65em #1}

\begin{document}
\title{The QCD phase transition with physical-mass, chiral quarks}

\collaboration{HotQCD Collaboration}
\newcommand\bnl{Physics Department, Brookhaven National Laboratory,Upton, NY 11973, USA}
\newcommand\llnl{Physics Division, Lawrence Livermore National Laboratory, Livermore CA 94550, USA}
\newcommand\cu{Physics Department, Columbia University, New York, NY 10027, USA}
\newcommand\bfld{Fakult\"at f\"ur Physik, Universit\"at Bielefeld, D-33615 Bielefeld, Germany}
\newcommand\bu{Center for Computational Science, Boston University, Boston, MA 02215, USA}
\newcommand\inth{Institute for Nuclear Theory, Box 351550, Seattle, WA 98195-1550, USA}
\newcommand\wh{Key Laboratory of Quark and Lepton Physics (MOE) and Institute of Particle Physics, Central China Normal University, Wuhan 430079, China}
\newcommand\lanl{Theoretical Division, Los Alamos National Laboratory, Los Alamos, NM 87544, USA}

\author{Tanmoy Bhattacharya}\affiliation{\lanl}
\author{Michael I.~Buchoff}\affiliation{\llnl}\affiliation{\inth}
\author{Norman H.~Christ}\affiliation{\cu}
\author{H.-T.~Ding}\affiliation{\wh}
\author{Rajan Gupta}\affiliation{\lanl}
\author{Chulwoo Jung}\affiliation{\bnl}
\author{F.~Karsch}\affiliation{\bnl}\affiliation{\bfld}
\author{Zhongjie Lin}\affiliation{\cu}
\author{R.~D.~Mawhinney}\affiliation{\cu}
\author{Greg McGlynn}\affiliation{\cu}
\author{Swagato Mukherjee}\affiliation{\bnl}
\author{David Murphy}\affiliation{\cu}
\author{P.~Petreczky}\affiliation{\bnl}
\author{Chris Schroeder}\affiliation{\llnl}
\author{R~A.~Soltz}\affiliation{\llnl}
\author{P.~M.~Vranas}\affiliation{\llnl}
\author{Hantao Yin}\affiliation{\cu}


\begin{abstract} 
We report on the first lattice calculation of the QCD phase transition using chiral fermions at physical values of the quark masses.  This calculation uses 2+1 quark flavors, spatial volumes between (4 fm$)^3$ and (11 fm$)^3$ and temperatures between 139 and 196 MeV .  Each temperature was calculated using a single lattice spacing corresponding to a temporal Euclidean extent of $N_t=8$.   The disconnected chiral susceptibility, $\chi_{\rm disc}$ shows a pronounced peak whose position and height depend sensitively on the quark mass.  We find no metastability in the region of the peak and a peak height which does not change when a 5 fm spatial extent is increased to 10 fm.  Each result is strong evidence that the QCD ``phase transition'' is not first order but a continuous cross-over for $m_\pi=135$~MeV.  The peak location determines a pseudo-critical temperature  $T_c = 155(1)(8)$~MeV.  Chiral $\sua$ symmetry is fully restored above 164 MeV, but anomalous $\ua$ symmetry breaking is non-zero above $T_c$ and vanishes as $T$ is increased to 196 MeV. 
\end{abstract}

\pacs{11.15.Ha, 12.38.Gc}
\preprint{BNL-103837-2014-JA, CU-TP-1205, INT-PUB-14-003,LLNL-JRNL-650194
}

\maketitle

As the temperature of the QCD vacuum is increased above the QCD energy scale $\Lambda_{\rm QCD} = 300$~MeV, asymptotic freedom implies that the vacuum breaking of chiral symmetry must disappear and the familiar chirally-asymmetric world of massive nucleons and light pseudo-Goldstone bosons must be replaced by an $\sua$ symmetric plasma of nearly massless up and down quarks and gluons.  Predicting, observing and characterizing this transition has been an experimental and theoretical goal since the 1980's.  General principles are consistent with this being either a first-order transition for sufficiently light pion mass or a second-order transition in the $O(4)$ universality class at zero pion mass with cross-over behavior for non-zero $m_\pi$.  While second order behavior is commonly expected, first-order behavior may be more likely if anomalous $\ua$ symmetry is partially restored at $T_c$ resulting in an effective $U_L(2)\times U_R(2)$ symmetry~\cite{Pisarski:1983ms,Pelissetto:2013hqa}.

The importance of the $\sua$ chiral symmetry of QCD for the phase transition has motivated the widespread use of staggered fermions in lattice studies of QCD thermodynamics because this formulation possesses one exact chiral symmetry at finite lattice spacing, broken only by the quark mass.  However, the flavor symmetry of the  staggered fermion formulation is complicated showing an $SU_L(4)\times SU_R(4)$ ``taste'' symmetry that is broken by lattice artifacts and made to resemble the physical $\sua$ symmetry by taking the square root of the Dirac determinant, a procedure believed to have a correct but subtle continuum limit for non-zero quark masses.

Because of these limitations, it is important to study these phenomena using a different fermion formulation, ideally one which supports the full $\sua$ chiral symmetry of QCD at finite lattice spacing.  It is such a study which we report here.  We use M\"obius domain wall fermions~\cite{Brower:2012vk}, a formulation in which the fermions are defined on a five-dimensional lattice.  The extent in the fifth dimension, $L_s=16$ or 24, making the calculation at least 16 to 24 times more costly.  However, the resulting theory possesses an accurate $\sua$ symmetry, broken only by the input quark mass and the highly suppressed mixing between the left and right four-dimensional boundaries, where the low-energy fermions propagate.  This residual chiral asymmetry is a short-distance phenomenon whose leading long-distance effect is to add a constant $\mres$ to each input quark mass, $m_q$, giving a total mass $\widetilde{m}_q = m_q+\mres$.  Here the residual mass $\mres \approx 3$~MeV.  Additional residual chiral symmetry breaking is $O(a^2)$ smaller~\cite{Blum:2000kn}.

Because of the computational cost of this formulation, the calculation reported here uses only one lattice spacing, $a$, at each temperature, corresponding to a single temporal extent of $N_t=8$.  The good agreement with experiment for $f_\pi$ and $f_K$ computed at our largest lattice spacing and a comparison of zero temperature results at our $T\approx 170$  lattice spacing with results from two smaller lattice spacings~\cite{Arthur:2012opa}, suggest discretization errors of $\approx 5\%$  in our results.   In contrast, the less costly staggered fermion calculations are performed  using $N_t=8$, 10, 12 and 16.  However, to make a controlled continuum extrapolation, the staggered fermion discretization errors are assumed to behave as $a^2$.  Potential non-linearities in the taste-breaking effects, which in zero-temperature staggered fermion calculations are handled using staggered chiral perturbation theory, are ignored because of the absence of a corresponding theory of finite-temperature taste breaking.

\section{Methods}

The present calculation with $m_\pi=135$ MeV and $32^3\times8$ and $64^3\times8$ volumes extends earlier domain wall fermion results with $m_\pi=200$~MeV and $16^3\times8$, $24^3\times8$ and $32^3\times8$ volumes~\cite{Bazavov:2012qja, Buchoff:2013nra}.  We use the same combination of the Iwasaki gauge action and dislocation suppressing determinant ratio (DSDR) exploited to reduce residual chiral symmetry breaking in these earlier studies.  However, to enable calculations at $m_\pi = 135$ MeV with available computing resources we have changed the Shamir domain wall formulation to M\"obius~\cite{Brower:2012vk}.  By choosing the M\"obius parameters $b$ and $c$ of Ref.~\cite{Brower:2012vk} so that $b-c=1$, we insure that our M\"obius Green's functions will agree at the 0.1\% level with those of Shamir evaluated at a much larger $L_s$.  Thus, our $m_\pi=200$ and 135~MeV calculations are equivalent, including all lattice artifacts, except for the intended reduction in quark mass.

\begin{table}[!htp]
	\caption{A summary of the $m_\pi=135$~MeV ensembles.  The units are MeV for the temperature $T$ and $10^{-5}/a$ for the masses $m_l$, $m_s$ and $\mres$. $N_\text{st}$, $N_\text{tot}$ and $N_\sigma$ label the number of independent streams, the total equilibrated time units and the number of sites in each spatial direction, respectively.}
\label{tab:parms}
\begin{tabular}{cccccccccc}
$T$ 	& $\beta$	&$N_\sigma$	&$L_s$	&$c$&$m_l$	&$m_s$	&$\mres$&$N_\text{st}$		
													& $N_\text{tot}$ \\
\hline\hline
139	&1.633	&32			&24		&1.5	&\multicolumn{1}{r}{22}
											&5960	&219(1)	&4	&\multicolumn{1}{r}{5768} \\
''	&''		&64			&''		&''	&'' 		&'' 		&'' 		&1	&\multicolumn{1}{r}{380}   \\
149	&1.671	&32			&16		&1.5	&\multicolumn{1}{r}{34}
											&5538	&175(1)	&4	&\multicolumn{1}{r}{7823} \\
''	&''		&64			&''		&''	&'' 		&'' 		&'' 		&3	&\multicolumn{1}{r}{2853} \\
154	&1.689	&32			&16		&1.5	&\multicolumn{1}{r}{75}
											&5376	&120(4)	&4	&\multicolumn{1}{r}{6108} \\
159	&1.707	&32			&16		&1.5	&112	&5230	&\multicolumn{1}{r}{91(1)}
															&3	&\multicolumn{1}{r}{8714} \\
''	&''		&64			&''		&''	&'' 		&'' 		&'' 		&2	&\multicolumn{1}{r}{3431} \\
164	&1.725	&32			&16		&1.5	&120	&5045	&\multicolumn{1}{r}{68(5)}
															&4	&\multicolumn{1}{r}{7149} \\
168	&1.740	&32			&16		&1.2	&126	&4907	&\multicolumn{1}{r}{57(1)}
															&2	&\multicolumn{1}{r}{5840} \\
''	&''		&64			&''		&''	&'' 		&'' 		&'' 		&1	&\multicolumn{1}{r}{1200} \\
177	&1.771	&32			&16		&1.0	&132	&4614	&\multicolumn{1}{r}{43(1)}
															&2	&\multicolumn{1}{r}{8467} \\
186	&1.801	&32			&16		&1.0	&133	&4345	&\multicolumn{1}{r}{26(1)}
															&2	&\multicolumn{1}{r}{10127} \\
195	&1.829	&32			&16		&0.9	&131	&4122	&\multicolumn{1}{r}{19(1)}
															&2	&\multicolumn{1}{r}{10124} \\
\hline
\end{tabular}
\end{table}

Table~\ref{tab:parms} lists the parameters for the $m_\pi=135$~MeV ensembles and the measured values for the residual mass.  At the lowest temperatures, more than 90\% of the quark mass is generated by residual chiral symmetry breaking. In addition to these 13 ensembles with $N_t = 8$, two calculations were performed at $T=0$ with space-time  volume $32^3\times 64$.  These used $\beta = 1.633$ (first reported here) and $\beta = 1.75$~\cite{Arthur:2012opa} and correspond to our $T=139$ MeV and $T \approx 170$ MeV when $N_t=8$.

The choices of quark masses and assigned temperatures given in Tab.~\ref{tab:parms} were estimated from earlier work ~\cite{Bazavov:2012qja, Arthur:2012opa}.  Results from the new zero temperature ensemble at $\beta=1.633$, obtained with the quark masses shown in Tab.~\ref{tab:parms}, are summarized in Tab.~\ref{tab:1.633} and provide a check of these estimates.  The resulting lattice spacing and pion mass are close to our targets while the kaon mass is lighter than expected, which may be unimportant for the quantities studied here.  Of special interest is a comparison of the residual mass for this value of $\beta$ given in Tabs.~\ref{tab:parms} and \ref{tab:1.633}.  The 1.1\% discrepancy is a measure of discretization error.  Likewise the comparison with experiment of $f_\pi$ and $f_K$ gives 6\% and 4\% errors, indicating the size of discretization effects.

\begin{table}[!htp]
\caption{Results at $\beta=1.633$ and $T=0$ (in lattice units and MeV) from 25 configurations separated by at least 20 time units.  We use $M_\Omega$ to fix the scale.  Also listed are the experimental values.}	
\label{tab:1.633}
\begin{tabular}{cccc}
			& $1/a$			& MeV		& Expt.(MeV) 	\\
\hline\hline
$m_\pi$		&0.1181(5)		& 129.2(5)		&  135		\\
$m_K$		&0.4230(5)		& 462.5(5)		&  495		\\
$m_\Omega$	&1.530(3)		& 1672.45		&1672.45		\\
$T=\frac{1}{8a}$	&0.125		& 136.7(3)		&  \textemdash\		\\
$f_\pi$		&0.1263(2)		& 138.1(2)		& 130.4		\\
$f_K$		&0.1483(4)		& 162.2(4)		& 156.1		\\
$\mres$		&0.00217(2)		& \textemdash\  	& \textemdash\  		\\
\hline
\end{tabular}
\end{table}

\section{Results}

\begin{figure}[htb]
\begin{center}
\includegraphics[width=\linewidth]{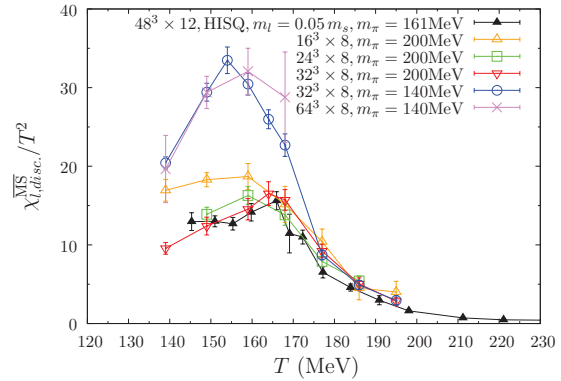}
\end{center} 
\caption{The dependence of the disconnected chiral susceptibility on temperature for $m_\pi=135$ and 200~MeV.  The $m_\pi=135$ MeV data shows a near $2\times$ increase over that for $m_\pi=200$~MeV.   HISQ results for $m_\pi=161$ MeV~\cite{Bazavov:2011nk, Buchoff:2013nra} are also plotted.}
  \label{fig:chi_disc}
\end{figure}

Our most dramatic result is the temperature-dependent, disconnected chiral susceptibility $\chi_{\rm disc}$, plotted in Fig.~\ref{fig:chi_disc}.   Three of the four lower curves show earlier results with $m_\pi=200$~MeV on $16^3$, $24^3$ and $32^3$ volumes.  A significant decrease in $\chi_{\rm disc}$ is seen for temperatures below 165~MeV as the volume is increased above $16^3$, a volume dependence anticipated in earlier scaling ~\cite{Engels:2001bq, Engels:2011km} and model~\cite{Braun:2010vd} studies.  The two higher curves show a large increase in $\chi_{\rm disc}$ in the entire transition region for $m_\pi=135$~MeV and both $32^3$ and $64^3$ volumes.   The ratio of peak heights for the $m_\pi=135$ and 200 MeV, $32^3$ data is 2.1(0.2), which is consistent with the ratio 1.86 predicted by universal $O(4)$ scaling $\sim \widetilde{m}_l^{1/\delta-1} \propto m_ \pi^{-1.5854}$, only if the regular, mass-independent part of $\chi_{\rm disc}$ is small. 

This comparison of $\chi_{\rm disc}$ with universal $O(4)$ scaling neglects the connected part of the chiral susceptibility.  We find that the connected chiral susceptibility has a mild dependence on both the temperature and quark mass (as is expected if the $\delta$ screening mass remains non-zero at $T_c$) and so does not contribute to the singular part of the chiral susceptibility.

Also shown in this figure are HISQ results for $N_t=12$ and a Goldstone pion mass of 161 MeV~\cite{Bazavov:2011nk, Buchoff:2013nra}.   If scaled to $m_\pi=135$ MeV assuming this same $m_ \pi^{-1.5854}$ behavior, the HISQ value for $\chi_{\rm disc}$  is 50\% smaller than that seen here.  This discrepancy reaffirms the importance of an independent study of the order of the transition and calculation of $T_c$ using chiral quarks.  

The peak shown in Fig.~\ref{fig:chi_disc} implies a pseudo-critical temperature of 155(1)(8)~MeV. The central value and statistical error are obtained by fitting the $T=149$, 154 and 159 MeV values of $\chi_{\rm disc}^{\overline{\rm MS}}$ to a parabola.  The second, systematic error reflects the expected 5\% discretization error.  We do not include a systematic error caused by our finite volume.  While typically neglected when $N_\sigma/N_t \ge 4$,  we lack the data needed for an empirical estimate.  This result for $T_c$ is consistent with the continuum limit for this quantity obtained using staggered fermions~\cite{Borsanyi:2010bp, Bazavov:2011nk}.  

\begin{figure}[htb]
\begin{center}
\includegraphics[width=\linewidth]{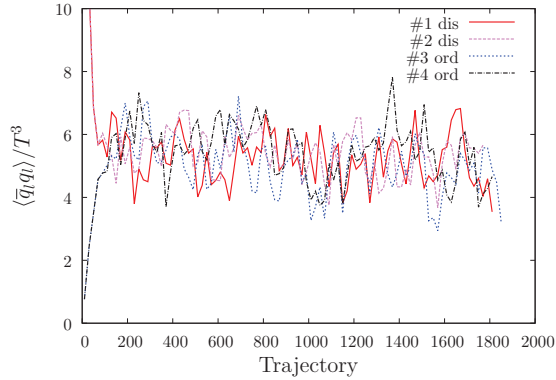}
\end{center} 
\caption{The time histories of $\langle\overline{q}_lq_l\rangle$ for four streams with $T=154$ MeV . Streams beginning with an ordered or disordered configuration are labeled ``ord'' and ``dis''.  Each point is the average of measurements made with 10 random sources on each of 20 configurations, separated by one time unit.} 
\label{fig:pbp_evol}
\end{figure}

The order of the QCD phase transition can now be studied using the time-history of the chiral condensate for $T \approx T_c$.  Figure~\ref{fig:pbp_evol} shows four time histories of $\langle\overline{q}_lq_l\rangle$ at $T=154$ MeV.  All four streams fluctuate over the same range of  values, showing no metastable behavior and no difference between those streams starting from ordered versus disorder configurations.  This and the failure of $\chi_{\mathrm{disc}}$ to grow as $2^3$ when the volume is increased from $32^3$ to $64^3$ provide strong evidence that  for $m_\pi = 135$ MeV, the QCD phase transition is not first-order but a cross-over, a conclusion consistent with previous staggered work~\cite{Bernard:2004je, Aoki:2006we, Cheng:2006qk,  Bazavov:2011nk}.  

\begin{figure}[htb]
\begin{center}
\includegraphics[width=0.93\linewidth]{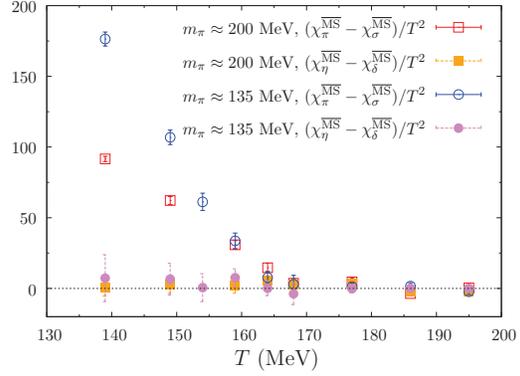}
\end{center} 
\caption{Two susceptibility differences are shown that reflect the $\sua$ symmetry of QCD and our chiral fermion formulation.  Below $T_c$ this symmetry is spontaneously broken.  For $T > 164$~MeV we see accurate chiral symmetry.}
\label{fig:sua}
\end{figure}

\begin{figure}[htb]
\begin{center}
\includegraphics[width=0.93\linewidth]{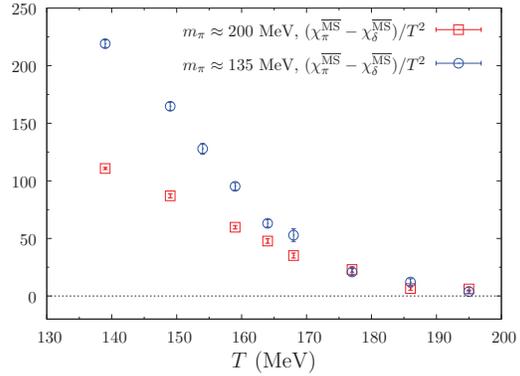}
\end{center} 
\caption{The $T$-dependence of the anomalous $\ua$-breaking difference $\chi_\pi-\chi_\delta$, which remains non-zero and becomes mass independent for $T > 168$ MeV.} 
\label{fig:ua}
\end{figure}

In Fig.~\ref{fig:sua} we show the $\sua$-breaking differences between the susceptibilities $\chi_\pi$ and $\chi_\sigma$ and between $\chi_\delta$ and $\chi_\eta$.  Each pair of fields, $(\vec \pi,\sigma)$ and $(\vec \delta,\eta)$ forms a 4-dimensional representation of $\sua$ symmetry.  These $\sua$-breaking differences are large below $T_c$ but have become zero for $T > 164$~MeV.  In Fig.~\ref{fig:ua} we show the difference $\chi_\pi-\chi_\delta$.   This pair of quantities is related by the anomalous $\ua$ transformation, a symmetry of the classical field theory that is broken by the axial anomaly.  Figure~\ref{fig:ua} shows that this symmetry is not restored until at least $T \ge 196$ MeV.  Also shown in this figure is the result from our earlier $m_\pi=200$~MeV calculation~\cite{Buchoff:2013nra}.  The expected increase in $\chi_\pi-\chi_\delta$ with decreasing pion mass is seen for $T \le T_c$.  However, above $T=168$ MeV this difference is still non-zero and has become mass independent, confirming our previous conclusion that this non-zero value is a result of the axial anomaly, not the small quark mass.

\section{Conclusion}

We have presented results from the first study of the QCD phase transition using chirally symmetric lattice fermions, physical quark masses and therefore three degenerate pions with $m_\pi \approx 135$ MeV.   We find $T_c=155(1)(8)$ MeV, similar to previous staggered fermion results, and see cross-over behavior, consistent with a second order critical point at zero quark mass.  We show that anomalous symmetry breaking extends to temperatures approximately 30~MeV above $T_c$.  Finally, we see a factor of two increase in the disconnected chiral susceptibility, $\chi_{\rm disc}$ near $T_c$ as $m_\pi$ decreases from 200 to 135 MeV, similar to the expectations for critical $O(4)$ scaling, provided the regular part of $\chi_{\rm disc}$  is small.  However, in this region we find $\chi_{\rm disc}$ 50\% larger than that suggested by staggered fermion results.

These results may close a chapter in the study of the QCD phase transition.  The cross-over character and pseudo-critical temperature of the transition have now been obtained using a formulation which respects the symmetries of QCD, uses physical values for both the strange and light quark masses and is performed for values of the inverse lattice spacing $1/a \ge 1.1$ GeV where  5\% discretization errors are to be expected. This is a challenging calculation with 5-dimensional lattice volumes as large as $64^3\times8\times 24$ and a physically light quark mass.  This study was made possible by the use of the DSDR action~\cite{Renfrew:2009wu}, M\"obius fermions~\cite{Brower:2012vk}, highly efficient code~\cite{Boyle:2012iy} and substantial resources provided by the Lawrence Livermore National Laboratory.  Of course, it remains important to continue to explore these questions at larger spatial volume and smaller lattice spacing when adequate resources become available.

\section{Acknowledgements}

Computing support for this work on the LLNL Vulcan BG/Q supercomputer came from the Lawrence Livermore National Laboratory (LLNL) Institutional Computing Grand Challenge program. We thank LLNL for funding from LDRD13-ERD-023. This work has been supported by the U. S. Department of Energy under contract DE-AC52-07NA27344 (LLNL).  NHC, ZL, RDM, GM, DM and  HY were supported in part by DOE grant DE-FG02-92ER40699. TB and RG were supported in part by DOE grant DE-KA-1401020.  H-TD, FK, SM and PP were supported in part by DOE grant DE-AC02-98CH10886.  MIB was supported in part by DOE grant DE-FG02-00ER41132 This work and the development of software used in this work has been supported in part through the Scientific Discovery through Advanced Computing (SciDAC) program funded by the U.S. Department of Energy, Office of Science, Advanced Scientific Computing Research and Nuclear Physics.

\end{document}